# FeTe as a candidate material for new iron-based superconductor

Y. Mizuguchi[a,b,c,*], F. Tomioka[a,b], S. Tsuda[a,b,d], T. Yamaguchi[a,b], Y. Takano[a,b,c]

[a] National Institute for Materials Science, 1-2-1, Sengen, Tsukuba, Ibaraki 305-0047, Japan

[b] JST, TRIP, 1-2-1, Sengen, Tsukuba, Ibaraki 305-0047, Japan

[c] University of Tsukuba, 1-1-1, Tennodai, Tsukuba, Ibaraki 305-0006, Japan

[d] WPI-MANA-NIMS, 1-1, Namiki, Tsukuba, Ibaraki 305-0044, Japan

Abstract

Tetragonal FeSe is a superconductor with a transition temperature $T_c$ of 8 K and shows a huge enhancement of $T_c$ with applying pressure. Tetragonal FeTe has a structure very analogous to superconducting FeSe, but does not show superconducting transition. We investigated the pressure effect of resistivity on FeTe. The resistivity at room temperature decreased with increasing pressure. An anomaly in resistivity around 80 K shifted towards a lower temperature with increasing pressure.

* Corresponding author.

Mr. Yoshikazu Mizuguchi

Postal address: National Institute for Materials Science, 1-2-1, Sengen, Tsukuba, Ibaraki 305-0047, Japan

Phone: +81-29-859-2514

Fax: +81-29-859-2601

E-mail address: MIZUGUCHI.Yoshikazu@nims.go.jp

1. Introduction

A great discovery of a new superconductor LaFeAsO$_{1-x}$F$_x$ with a transition temperature $T_c$ of 26 K [1] triggered active studies on iron-based superconductors. The $T_c$ was raised to 43 K by applying pressure [2]. The $T_c$ increased above 50 K with substitution of smaller ion for La [3-6]. Related iron-based superconductors are Ba$_{1-x}$K$_x$Fe$_2$As$_2$ [7] and Li$_{1-x}$FeAs [8], which have FeAs layers analogous to LaFeAsO$_{1-x}$F$_x$. Recently, superconductivity at 8 K in tetragonal FeSe was reported [9]. X-ray and neutron diffraction patterns indicated a structural phase transition from tetragonal to orthorhombic around 70 K [10]. In $^{77}$Se NMR measurement, the nuclear-spin lattice relaxation ratio $1/T_1$ showed $T^3$ behavior below $T_c$ without a coherence peak [11]. The similar behaviors have been reported in some of the iron-based superconductors. We reported an extreme enhancement of $T_c$ under high pressure [12]. The onset of $T_c$ increased from 13 to 27 K at 1.48 GPa. The enhancement of $T_c$ by the substitution of Te for Se was reported [13,14]. Density functional calculations for FeS, FeSe and FeTe indicated a strong stability of spin density wave for FeTe and predicted a possibility of superconductivity with higher $T_c$ in doped FeTe than FeSe [15].

Tetragonal FeTe has a structure very analogous to superconducting FeSe.

However, FeTe dose not show superconductivity. FeTe shows an anomaly, corresponding to a structural transition, in resistivity around 80 K. If the structural transition can be suppressed, superconductivity may appear in FeTe, as in other iron-based superconductors. Here we report a structural analysis and a pressure effect on the resistivity of FeTe$_{0.92}$.

2. Experiments

FeTe$_{0.92}$ sample was prepared using a solid state reaction method. Fe powders and Te shots were sealed into an evacuated quartz tube with a nominal composition of Fe:Te = 1:0.92 and heated at 800 K for 12 h. The obtained samples contained single crystals. We well ground the crystals and characterized the samples by X-ray diffraction using CuK$_\alpha$ radiation. The X-ray diffraction profile was refined using Rietan2000.

Resistivity measurements were performed using a four terminal method from 300 K to 2 K. The connections to the measuring device were made by using gold wires of 20 μm in diameter attached to the sample by a silver paste. Hydrostatic pressures were generated by a BeCu/NiCrAl clamped piston-cylinder cell. The sample was immersed in a fluid pressure-transmitting medium of Fluorinert (FC70:FC77 = 1:1) in a Teflon cell. The actual pressure was estimated by a superconducting transition temperature of Pb. Resistivity at 300 K was normalized using resistivity of a bulk sample of FeTe$_{0.92}$.

3. Results and discussions

Figure 1 shows the X-ray diffraction profile and a result of Rietveld refinement. The X-ray profile of FeTe$_{0.92}$ confirmed the primitive tetragonal unit cell ($a$ = 3.8248(3) Å, $c$ = 6.2910(3) Å, $V$ = 92.03(1) Å$^3$, space group $P4/nmm$, $R_{wp}$ = 10.11%, $z$-coordination parameter = 0.2790(2)). The Te content refined to a value of 0.937(10). A Te-Fe-Te angle ($\alpha$-angle) was estimated to be ~95º. Figure 2 shows the crystal structure of FeTe$_{0.92}$.

Figure 3 and 4 show a temperature dependence of resistivity for FeTe$_{0.92}$ under high pressures and an enlargement of below 140 K. At ambient pressure, the resistivity increased with decreasing temperature down to ~100 K. We observed a hump, which corresponds to a structural transition from tetragonal to orthorhombic [14,16], around a transition temperature $T_s$ = 82.7 K. $T_s$ was estimated to be a temperature with a maximum point of $d\rho/dT$. Below $T_s$, the temperature dependence of resistivity showed metallic behavior. Superconductivity was not observed down to 2 K.

With increasing pressure, the resistivity decreased drastically. We estimated $T_s$

at 0.45, 1.07 and 1.60 GPa to be 79.8, 70.4 and 53.1 K. The obtained $T_s$ was plotted in Fig. 5 as a function of applied pressure. The $T_s$ evidently decreases with increasing pressure. However, contrary to our expectation, superconductivity has not been observed within current pressure region. How can we induce superconductivity in $FeTe_{0.92}$? We expect that further pressure will induce superconductivity for $FeTe_{0.92}$.

The estimated α-angle of $FeTe_{0.92}$ (α = 95º) is significantly smaller than that of superconducting $FeSe_x$ (α = 104.54º, $x \sim 0.08$). The α-angle is likely to be important for determination of $T_c$ in FeAs-based superconductor [17]. The α-angle near 104º may be necessary for pressure-induced superconductivity in $FeTe_{0.92}$.

4. Summary

We synthesized a single phase of $FeTe_{0.92}$ and performed resistivity measurement under high pressure. We observed a shift of $T_s$ to a lower temperature with increasing pressure. Our results suggest a possibility of pressure-induced superconductivity in $FeTe_{0.92}$ with higher pressure than 1.60 GPa.

Acknowledgement

This work was partly supported by Grant-in-Aid for Scientific Research

Matsuhata, M. Braden, K. Yamada, *J. Phys. Soc. Jpn.* 77 (2008) 083704.

Figure captions

Fig. 1. X-ray diffraction pattern of FeTe$_{0.92}$ and a result of Rietveld refinement.

Fig. 2. Crystal structure of FeTe0.92. The graphics were drawn using VESTA.

Fig. 3. Temperature dependence of resistivity for FeTe$_{0.92}$ under high pressure. An arrow indicates the estimated $T_s$.

Fig. 4. Enlargement of resistivity below 140 K. An arrows indicate the estimated $T_s$

Fig. 5. Pressure dependence of $T_s$.

Fig.1

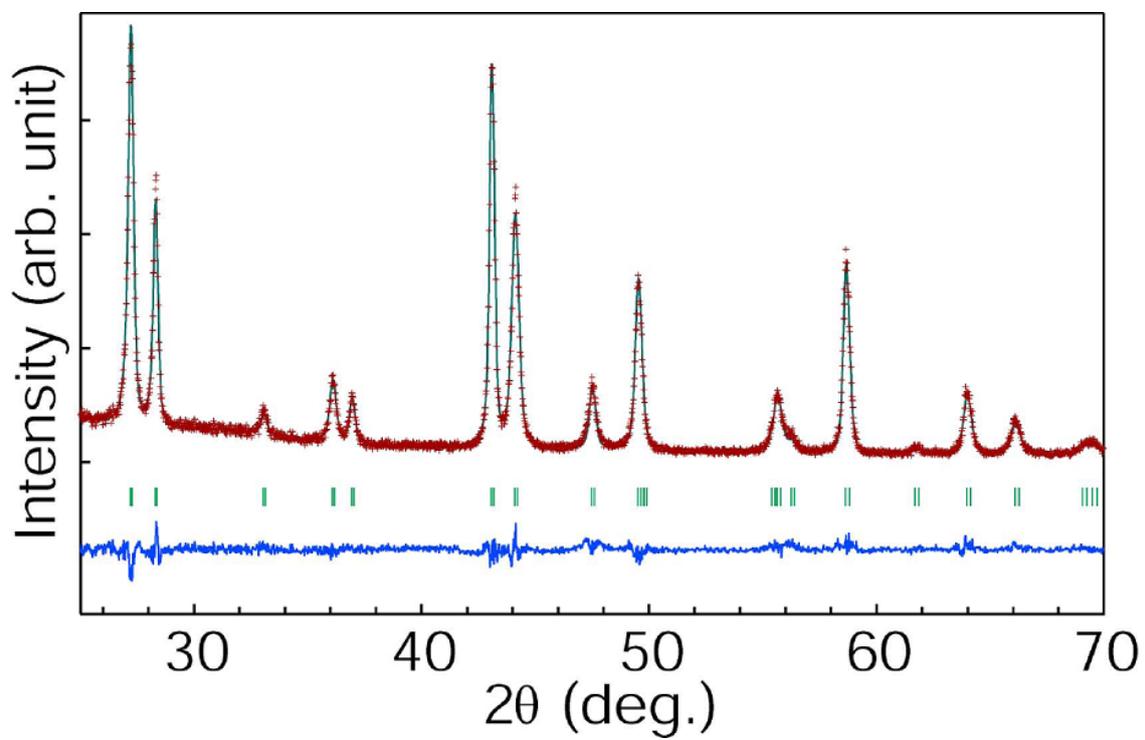

Fig. 2

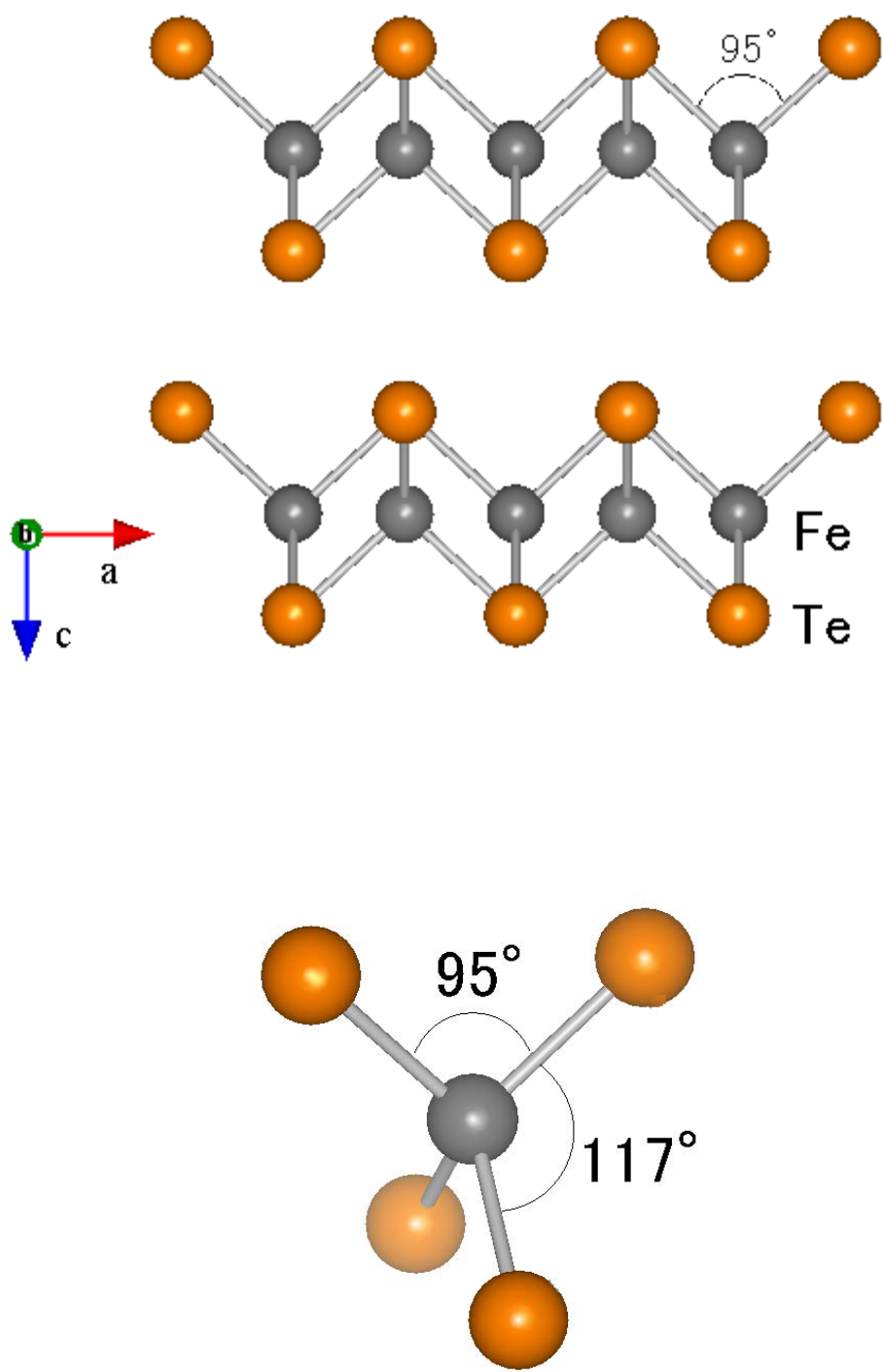

Fig. 3

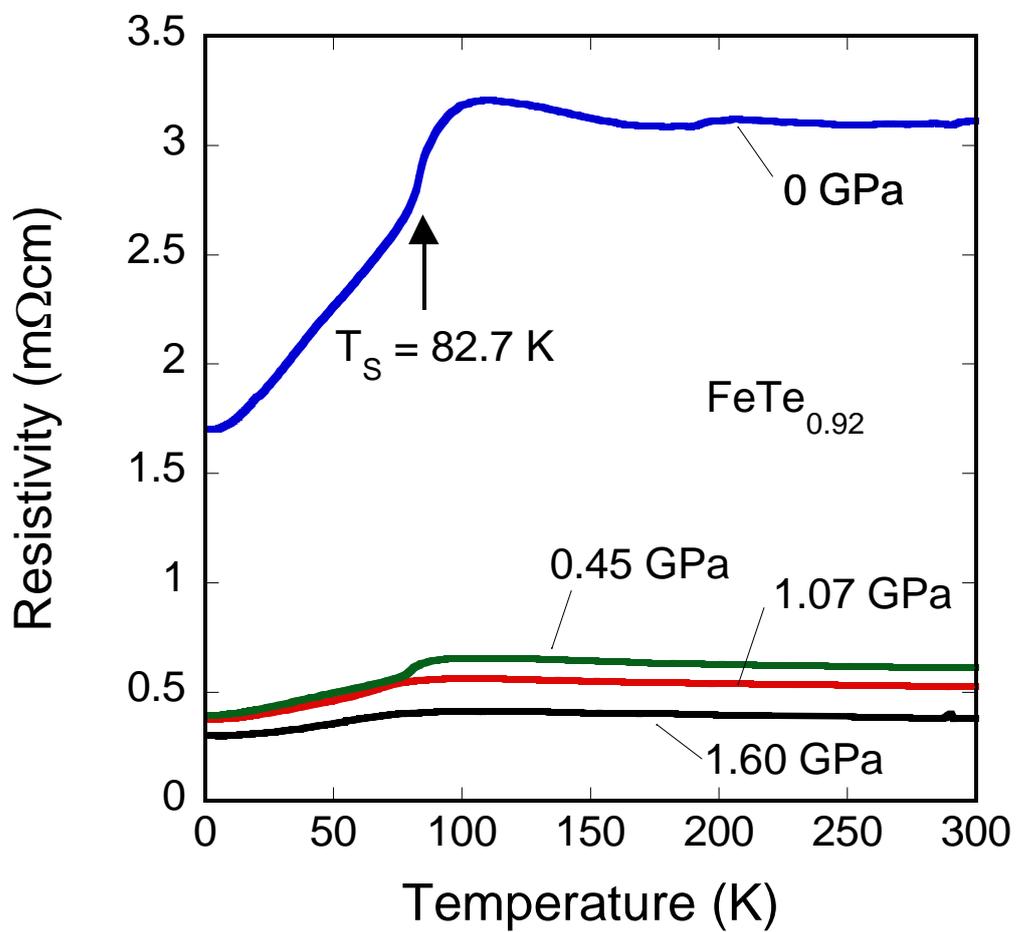

Fig. 4

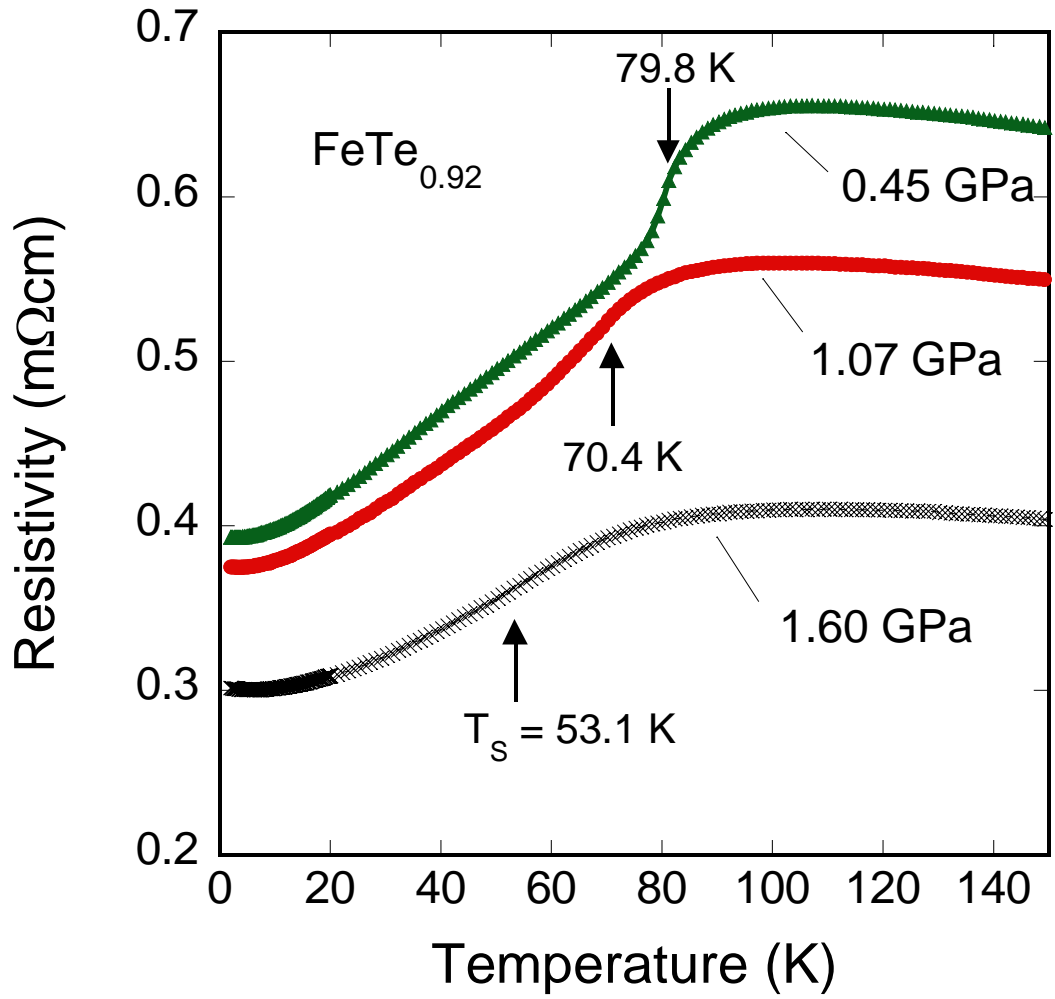

Fig. 5

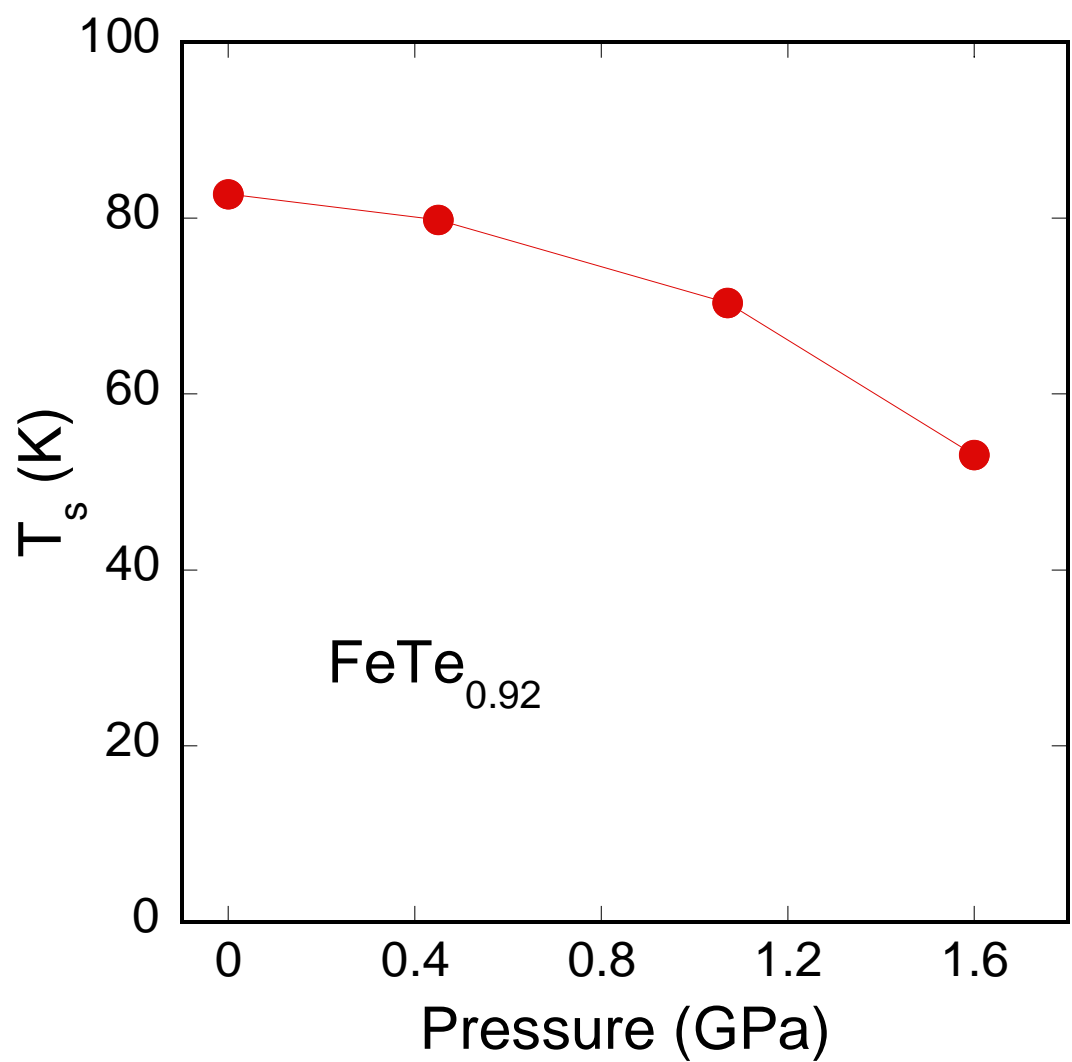